# Fully Complex Magnetoencephalography

*Jonathan Z. Simon*[a,b,c] *and Yadong Wang*[c]

[a]Department of Electrical & Computer Engineering
[b]Department of Biology
[c]Program in Neuroscience and Cognitive Science
University of Maryland, College Park, MD, USA

*Pages*: 23 (including figure captions and table)
*Tables*: 1
*Figures*: 5

*Corresponding Author:*
Jonathan Z. Simon

Electrical & Computer Engineering Department
University of Maryland
College Park MD 20742

Tel: 1-301-405-3645
Fax: 1-301-314-9281
E-mail address: jzsimon@eng.umd.edu (J. Z. Simon)

**Abstract**

Complex numbers appear naturally in biology whenever a system can be analyzed in the frequency domain, such as physiological data from magnetoencephalography (MEG). For example, the MEG steady state response to a modulated auditory stimulus generates a complex magnetic field for each MEG channel, equal to the Fourier transform at the stimulus modulation frequency. The complex nature of these data sets, often not taken advantage of, is fully exploited here with new methods. Whole-head, complex magnetic data can be used to estimate complex neural current sources, and standard methods of source estimation naturally generalize for complex sources. We show that a general complex neural vector source is described by its location, magnitude, and direction, but also by a phase and by an additional perpendicular component. We give natural interpretations of all the parameters for the complex equivalent-current dipole by linking them to the underlying neurophysiology. We demonstrate complex magnetic fields, and their equivalent fully complex current sources, with both simulations and experimental data.

**1. Introduction**

Physiological questions of the human brain that demand temporal resolution commensurate with neuronal activity require electromagnetic techniques, particularly electroencephalography (EEG) (see, e.g. Gevins et al., 1995) or magnetoencephalography (MEG) (see, e.g. Hari and Lounasmaa, 1989; Lounasmaa et al., 1996). A compelling advantage of MEG is that it allows simultaneous spatial localization ("imaging") and high temporal resolution physiology of the neural sources (Roberts et al., 2000; Krumbholz et al., 2003). Neural sources' ionic currents generate measurable magnetic fields according to the classical physical equations of electrodynamics. The small magnetic signals (hundreds of femtoteslas) propagate outward transparently and can be measured with superconducting quantum interference devices (SQUIDs) (Hamalainen et al., 1993). The types of MEG responses whose source location and stimulus-related properties are commonly interpreted include evoked fields at specific latencies, e.g. the auditory N100 response (Hari et al., 2000) or evoked high frequency responses (Hashimoto et al., 1996); evoked or induced oscillatory responses (Hari and Salmelin, 1997; Lin et al., 2004); and steady state responses (SSR) to ongoing stimuli (Ross et al., 2000). SSR responses are a rich source of neurophysiological data but have received comparatively less attention.

Complex numbers arise naturally whenever any data, such as that from MEG, are analyzed with the Fourier transform. The Fourier transform takes a real valued time-varying signal and represents the same signal by a complex valued function of frequency. The original signal, at a one time instant, is represented by a single real number, but the Fourier transform, for a particular frequency, is represented by two real numbers, e.g. a magnitude and a phase. The magnitude is a non-negative number, and the phase is an abstract angle that varies from $0°$ to $360°$ (equivalently, $2\pi$ radians, or 1 cycle). Just as real numbers can be usefully generalized to complex numbers, real valued fields can be generalized to complex valued fields, and, in particular, real valued vector fields can be generalized to complex valued vector fields. In the case of MEG signals, the Fourier transform of the time varying magnetic field generates a complex valued magnetic field, for every spatial point (channel) the field is measured. Related transforms, such as wavelet and other short time Fourier transforms, also result in complex valued magnetic fields.

The utility of these complex valued responses can especially be seen in experiments and analysis that use SSR paradigms. In such paradigms, a stationary stimulus with periodic structure generates a neural response with the same periodic structure. Example auditory stimuli include: narrow- or broad-band carriers with periodically modulated amplitude, and periodic trains of clicks or tone-pips. In each case, there is a corresponding neural response with the same periodicity. The MEG SSR for sinusoidally amplitude modulated tones has been well documented (Ross et al., 2000; Ross et al., 2002; Schoonhoven et al., 2003) and the SSR in EEG has a long and rich history (Galambos et al., 1981). The strongest frequency response is at the stimulus modulation frequency (harmonic responses are substantially weaker and so are not treated here directly, though their generalizations are straightforward). The response at the modulation frequency gives a complex magnetic field: a magnetic field with amplitude as well as phase as information.

The amplitude simply gives the strength of the response at the modulation frequency. The phase corresponds to the time-delay of the response in units of the modulation frequency, when the phase is measured in cycles. Thus, a 0.010 s delay for a 10 Hz modulation frequency gives a phase of 0.1 cycles ($36°$, or $0.2\pi$ radians). The periodicity property of phase arises from the inability to distinguish time shifts longer than one cycle from the equivalent time shifts shorter than one cycle.

Beyond this simple interpretation, however, the complex nature of these data is not often exploited (some statistical techniques used in EEG do embrace the complex nature of the response, e.g. Picton et al., 2001; Picton et al., 2003). A simple example is the spatial distribution of phase over the whole head. Multi-channel MEG and EEG data is known for difficulty in its visualizability due to high dimensionality: many channels, many experimental conditions, and many repetitions, each a function of time. A greatly simplified picture results from replacing, for each channel, the entire dimension of time with the single value of the phase (of the frequency of interest). This representation has been used for EEG data analysis (Herdman et al., 2002). Examples of MEG whole head complex fields in response to auditory stimuli are shown in Fig. 1. In each case, the complex whole head SSR can be analyzed visually at once, whereas the comparable whole head response in the time domain (a time waveform displayed over every sensor) is difficult to absorb visually.

The utility of the complex nature of the data goes beyond the field distribution. A complex magnetic field is generated by its complex neural current source, a concept that has only been partially exploited in analysis of data from MEG (Lutkenhoner, 1992) and EEG (Lehmann and Michel, 1989; 1990; Michel et al., 1992).

Several approaches are typically used in MEG analysis to determine the neural current sources of a measured magnetic field (Baillet et al., 2001). One of the simplest is the equivalent-current dipole approximation, which uses a least-squares minimization algorithm, plus simplifications of the physics due to Sarvas (1987). The result of this method is a set of equivalent-current source dipoles. When applied to real magnetic field configurations, the resulting equivalent-current dipoles are real. A real equivalent-current dipole is defined by its location and a real dipole vector $\mathbf{q}$. Three real numbers are needed to fully describe a real vector: the three Cartesian components $(q_x, q_y, q_z)$, or equivalently, a two-dimensional orientation $(\theta,\phi)$ and an intensity ($q$).

A complex magnetic field configuration leads to complex equivalent-current dipoles, each of which, in addition to its location, is described by three complex numbers, or equivalently six real numbers. These can be seen as three complex components, or equivalently the six numbers given by the real and imaginary parts of the three Cartesian components $(\mathrm{Re}\{q_x\}, \mathrm{Re}\{q_y\}, \mathrm{Re}\{q_z\}, \mathrm{Im}\{q_x\}, \mathrm{Im}\{q_y\}, \mathrm{Im}\{q_z\})$. One may attempt to describe a complex dipole vector solely by its orientation (two real numbers) and a complex generalization of the intensity (two real numbers, e.g. a magnitude and phase), but this does not cover all six degrees of freedom. Nevertheless, a generic, complex, equivalent-current dipole can be described naturally and physiologically, in such a way that four of the six degrees of freedom do correspond to orientation and a complex intensity, and the two others are described below.

We discuss the roles and properties of the complex magnetic fields measured by MEG and SSR, which naturally lead to a visualization tool, the “whole-head complex SSR”.

The inverse problem is solved for a complex magnetic field distribution by determining the complex equivalent-current dipoles. The properties of complex dipoles are described, including all six degrees of freedom. Simulations are shown, and the method's utility is demonstrated with an example of a transfer function computation and an analysis of the variability of neural sources as a function of stimulus parameters.

The general methods outlined here are not special to MEG. Only small modifications are necessary to apply several of these methods to EEG and related techniques.

## 2. Methods

### *2.1. Complex Magnetic Fields from MEG and SSR Analysis*

A whole-head map of complex SSR responses is obtained by Fourier transforming each channel's response and focusing on the stimulus modulation frequency. For a stimulus with modulation frequency $f_{mod}$ and response measurement duration *T*, and an integer multiple of the cycle period ($T = N_{cyc} / f_{mod}$), the SSR complex response is component $N_{cyc}$ of the discrete Fourier transform of the response time waveform (the DC response is component zero). We assume that the MEG sensors are simple (not vector) magnetometers or gradiometers, giving one sampled time-waveform per channel.

One whole-head response pattern is shown in Fig. 1a. Each sensor's complex response is depicted by a "phasor", an arrow whose magnitude is proportional to the response magnitude and whose direction corresponds to the phase. The phase convention used here is the standard Cartesian convention: $0°$ phasors point to the right, and increasing phase corresponds to counterclockwise rotation. Fig. 1b-e shows the whole head SSR for four separate modulation frequencies.

[Fig. 1 about here]

The whole head complex SSR can optionally add magnetic field contours by projecting the complex values onto a line in the complex plane of constant phase: the complex numbers are turned into real numbers by rotating them by the line's phase and then taking the real part. This visual aid can greatly increase a viewer's ability to see natural structures, such as dipolar configurations. To underscore this, the inset of Fig. 1a shows the whole head complex SSR without the magnetic field contours, and the dipolar patterns are substantially more difficult to see (compared to the otherwise identical graphic in Fig. 1c). The line's phase can be chosen in several ways, but one method is to use the phase of the maximal spatial variance as measured over half the modulation cycle. This results in strong peaks (or troughs) of the projection whenever the phases strongly coincide (or anti-coincide) with that phase giving the most typical strong response. Only half the modulation cycle is used since the variance of a periodic signal has two peaks over an entire cycle.

There is an unavoidable ambiguity that a line with any particular phase is the same as that with the same phase plus 180°, which is equivalent to swapping positive and negative values of the projected field values. For auditory responses, this ambiguity can be often fixed by choosing a particular convention, e.g. that the positive/negative projected distribution has the same overlay of that of the source/sink distribution of a classic M100 response.

Another ambiguity that has been fixed is how the *phasor directions* correspond to phase. This ambiguity is important because of the unavoidable feature in this visual representation that directions on the printed page correspond to anatomical directions (i.e. the sensor layout) and, independently, to phase angles. The Cartesian coordinates used for the phasors in Fig. 1 are standard but arbitrary, and they may imply a vector flow where none exists. For instance in Fig. 1a, there appears to be a medial and posterior flow from the right frontal quadrant. This is entirely an artifact, and if the phases were plotted with the standard compass convention (0° upward and increasing phase rotates clockwise), the visual impression would instead be a divergence.

## *2.2. The Complex Equivalent-Current Dipole*

### 2.2.1. The Complex Inverse Problem

A commonly used technique that determines neural current sources from their generated magnetic field data can be straightforwardly generalized to complex fields. The resulting neural current source is a complex equivalent-current dipole (Lutkenhoner, 1992).

For example, the forward problem (the magnetic field due to a current dipole source) uses the complex version of the spherical head model (Sarvas, 1987; Mosher et al., 1999): outside a spherical conductor, the *complex* magnetic field **b** at a sensor with location **r** is generated by a *complex* current dipole **q** at location $\mathbf{r}_q$. The complex magnetic field due to multiple current dipoles is the linear sum of the multiple contributions. Since the complex magnetic field is linear with respect to the complex dipole moment $\mathbf{q}$ and nonlinear with respect to the location $\mathbf{r_q}$, we can generalize the linear model of the first stage of the inverse problem (Baillet et al., 2001) to complex quantities. For measurements made at *N* sensors by *p* dipoles, we can obtain $\mathbf{M} = \mathbf{A}\mathbf{S}^T$, where **M** is a columnar array of complex magnetic field measurements, $\mathbf{S}^T$ is a columnar array of complex dipole strengths, and **A**, the lead field matrix, is implicitly defined by the linear relationship between **b** and **q**, and is always real. In the presence of measurement errors, the model may be represented as $\mathbf{M} = \mathbf{A}\mathbf{S}^T + \varepsilon$, for $\varepsilon$ a complex error matrix. The least-squares (LS) method defines a cost function to minimize,

$$J_{LS} = \left\| \mathbf{M} - \mathbf{A}\mathbf{S}^{\mathbf{T}} \right\|_F^2 , \tag{1}$$

the Frobenius norm of the complex error matrix. For any set of sensor locations and complex dipole locations, the resulting array of complex dipole strengths, $\mathbf{S}^T$, is the one that minimizes $J_{LS}$, i.e. $\mathbf{S}^T = \mathbf{A}^{+}\mathbf{M}$, for $\mathbf{A}^{+}$ the pseudoinverse of $\mathbf{A}$. Lastly, the dipole location is obtained by minimizing $J_{LS}$. Minimization methods range from grid search and downhill simplex searches to global optimization schemes (Uutela et al., 1998).

It should be emphasized that the key feature of this method is the generalization of both the magnetic field and the source vectors to complex quantities (Lutkenhoner, 1992). Aside from this essential difference, the algorithm is unchanged from the real version.

Related algorithms that estimate a vector neural source (or source distribution) can be generalized analogously.

### 2.2.2. THE COMPLEX VECTORS

Like its real counterpart, the complex equivalent-current dipole is described by a location and a vector, but in this case, the vector is complex, with twice the degrees of freedom of a real vector. Any complex vector can be decomposed into its real and imaginary components, each a vector itself,

$$\mathbf{v} = \mathbf{v}_{\mathbf{Re}} + j\,\mathbf{v}_{\mathbf{Im}} . \qquad (2)$$

Its magnitude is given by the sum of its component magnitudes, $|\mathbf{v}|^2 = |\mathbf{v}_{\mathbf{Re}}|^2 + |\mathbf{v}_{\mathbf{Im}}|^2$ . A real vector has 3 degrees of freedom: a spatial orientation (two degrees of freedom) and a length (one degree of freedom), so a complex vector has six degrees of freedom.

From the complex vector, it is convenient to define a phase-parameterized real vector

$$\mathbf{v}(\theta) = \mathbf{v}_{\mathbf{Re}} \cos(\theta) + \mathbf{v}_{\mathbf{Im}} \sin(\theta) \qquad (3)$$

which defines the family of vectors swept out over the course of one cycle. The swept curve is an ellipse; an example is illustrated in Fig. 2.

[Fig. 2 about here]

At the start of the cycle, the vector is given entirely by its real component vector $\mathbf{v}_{\mathbf{Re}}$. As $\theta$ moves through the cycle, the vector mixes $\mathbf{v}_{\mathbf{Re}}$ and $\mathbf{v}_{\mathbf{Im}}$, until by $\theta = 90°$ the vector is given by $\mathbf{v}_{\mathbf{Im}}$. Note that, as shown in Fig. 2, the phase $\theta$ does not correspond to a spatial angle, since $\mathbf{v}_{\mathbf{Re}}$ and $\mathbf{v}_{\mathbf{Im}}$ are separated by $90°$ of phase but are not in general perpendicular. An ellipse can also be characterized by its semimajor and semiminor axes, $\mathbf{v}_{\mathbf{Max}}$ and $\mathbf{v}_{\mathbf{Min}}$, which are the swept vectors when $\mathbf{v}(\theta)$ reaches its maximum and minimum magnitudes, i.e. at the phases $\theta_{Max}$ and $\theta_{Min}$. It can be shown that

$$\theta_{Max} = \tfrac{1}{2}\left(\arg\left[-2\,\mathbf{v}_{\mathbf{Re}} \cdot \mathbf{v}_{\mathbf{Im}} - j\left(|\mathbf{v}_{\mathbf{Re}}|^2 - |\mathbf{v}_{\mathbf{Im}}|^2\right)\right] + \pi\right) \qquad (4)$$

and $\theta_{Min} = \theta_{Max} + 90°$. In the special case of the difference between $\theta_{Max}$ and $\theta_{Min}$, a phase advance of $90°$ does correspond to a spatial angle of $90°$ since semimajor and semiminor axes are always spatially perpendicular. Note also that a particular orientation with phase $\theta_{Max}$ is physically indistinguishable from the opposite orientation and $\theta_{Max} + 180°$. This ambiguity can be fixed by always requiring $0 \le \theta_{Max} < 180°$, but other resolutions may be more appropriate, e.g. unwrapping $\theta_{Max}$ smoothly for small stimulus parameter changes, which is the method used below.

Another useful parameter of the complex vector is its sharpness $\eta$, where

$$\eta = \frac{|\mathbf{v}_{\mathbf{Min}}|}{|\mathbf{v}_{\mathbf{Max}}|} \tag{5}$$

and $0 \le \eta \le 1$. When $\eta \approx 0$, the ellipse is highly elongated (very sharp, or eccentric) along the axis parallel to $\mathbf{v}_{\mathbf{Max}}$. Conversely, when $\eta = 1$, the ellipse degenerates into a circle. The sharpness $\eta$ is related to the eccentricity of the ellipse, *e*, by $e^2 = 1 - \eta^2$.

#### 2.2.3. Single Orientation Approximation

For complex dipole vectors whose swept ellipse is very sharp, the complex dipole vector simplifies. In the limit $\eta = 0$, the path simplifies to a straight line segment, whose ends are reached at the phases $\theta_{Max}$ and $\theta_{Max} + 180°$. This dipole can be described as having one orientation (the direction of $\mathbf{v}_{\mathbf{Max}}$), one strength ($|\mathbf{v}_{\mathbf{Max}}|$), and one phase ($\theta_{Max}$). The degree to which this is a good approximation is quantified by the sharpness $\eta$. This is a simpler generalization of a real vector than a general complex vector, adding only one degree of freedom (the phase) to the three degrees of freedom of a real vector.

When $\eta \ne 0$, we can still characterize a fully complex dipole vector by these same four degrees of freedom, but two extra degrees of freedom are needed: the sharpness $\eta$, and a second orientation, given by the azimuthal angle of the direction of $\mathbf{v}_{\mathbf{Min}}$ relative to the direction of $\mathbf{v}_{\mathbf{Max}}$. These two extra degrees of freedom bring the total to six, e.g. the three degrees of freedom of the real vector components plus three more from the imaginary components. In the special case of the Sarvas Model, the direction of the secondary orientation is constrained, since it must be orthogonal to both $\mathbf{v}_{\mathbf{Max}}$ and the radial direc-tion, and the only freedom left is whether the vector cross-product $\mathbf{v}_{\mathbf{Max}} \times \mathbf{v}_{\mathbf{Min}}$ (which must be perpendicular to both and therefore radial), is radially outward or inward.

*2.3. The Physiology of Complex Equivalent-Current Dipoles*

Recall that a real equivalent-current dipole is an *effective* (averaged) neural source: all the neural currents contributing to the measured magnetic field can be effectively replaced by one idealized source (Lutkenhoner, 2003). If the true source is compact, then the equivalent-current dipole is a good approximation of the location, strength, and orientation of the current source. Alternatively, if the true source is extended, then the equivalent-current dipole represents the averaged location, strength, and orientation of the extended neural source. In particular, there may be several distinct locations of neural sources, each with its own strength and orientation, but only the averaged quantities are expressed by the equivalent-current dipole.

In cases where the complex equivalent-current dipole vector's swept-out trajectory is approximately line-like, $\eta \approx 0$, the physiological interpretation is closely related to that of a real equivalent-current dipole but with one additional parameter, the phase. A complex dipole with high eccentricity oscillates at a single orientation; its phase corresponds to the delay, measured in cycles, of the oscillation's maximum. Indeed, an oscillating compact neural source can be described in entirety by its orientation, the phase at which the current is maximum, and the value of the maximum current.

A complex dipole with non-zero $\eta$ describes an effective source comprising an extended or distributed neural source(s): in this case more than one orientation, and its new corresponding strength, will be seen. For instance, several distinct neural sources in separate but nearby areas, with different strengths, orientations, and phases, will combine into a single complex equivalent-current dipole. The location of the single complex equivalent-current dipole will be an average of the locations of the distinct neural sources. The different strengths, orientations, and phases will average into *two* effective strengths and orientations and an overall phase ($\mathbf{v}_{\mathbf{Max}}, \mathbf{v}_{\mathbf{Min}}$, and $\theta_{Max}$). Or equivalently but more specifically, into a primary orientation and strength ($\mathbf{v}_{\mathbf{Max}}$), its phase ($\theta_{Max}$), the relative intensity in the direction of a secondary orientation ($\eta$), and the secondary orientation itself (described by a single azimuthal angle since it must also be perpendicular to $\mathbf{v}_{\mathbf{Max}}$).

Thus, sharpness can serve as an experimental measure of the extended or distributed nature of a neural source. A complex dipole of high $\eta$ is inconsistent with a single, compact neural source, and so indicates an extended source or multiple sources. $\eta$ near zero is consistent with a single, compact neural source and so is less likely to be generated by multiple sources.

*2.4. Evaluation of Neural Source Estimates*

Common evaluation techniques that measure how well the fitted data $\mathbf{M}_{fit}$ match the measured data $\mathbf{M}_{exp}$ also generalize to complex data. The correlation coefficient becomes complex and is given by

$$r = \frac{N\sum_{n=1}^{N} M_{fit,n}^{*} M_{exp,n} - \sum_{n=1}^{N} M_{fit,n}^{*} \sum_{n=1}^{N} M_{exp,n}}{\sqrt{\left(N\sum_{n=1}^{N}\left|M_{fit,n}\right|^{2} - \left|\sum_{n=1}^{N} M_{fit,n}\right|^{2}\right)\left(N\sum_{n=1}^{N}\left|M_{exp,n}\right|^{2} - \left|\sum_{n=1}^{N} M_{exp,n}\right|^{2}\right)}}, \quad (6)$$

where * is the complex conjugate operator. The phase of *r* expresses how much phase rotation should be applied to the fitted data to get a purely real *r* such that 0 < *r* < 1. The magnitude $|r|$ is what the value of *r* would be if the above rotation were applied, and has the same interpretation as for real *r* restricted to positive values. As in the real case, a perfect fit corresponds to $r = 1$, a fit that is otherwise perfect, except that the orientation is exactly opposite, corresponds to $r = -1$, and less-then perfect fits give $|r| < 1$. The complex case, however, allows additional phase offsets between the fitted and measured data.

The goodness of fit, being a power ratio, remains real and is given by

$$\mathrm{GOF} = 1 - \frac{\sum_{n=1}^{N} \left| M_{fit,n} - M_{exp,n} \right|^2}{\sum_{n=1}^{N} \left| M_{exp,n} \right|^2} \qquad (7)$$

where a GOF of 1 is a perfect fit. The main caveat for the GOF of complex distributions is that typical values are often much lower than for comparable real distributions. This is because complex distributions have twice as many degrees of freedom as real distributions (for the same number of channels), and the GOF distribution depends the number of degrees of freedom (c.f. the statistical *F* distribution).

### *2.6. Auditory MEG SSR Experimental Methods*

Sinusoidally amplitude-modulated sounds of 1 s duration were presented to three subjects (two male). The 12 stimuli had four modulation frequencies (16 Hz, 32 Hz, 48 Hz and 64 Hz) and three carriers (pure tone; 1/3 octave pink noise; 2 octave pink noise; all centered at 400 Hz). All 12 stimuli were presented 100 times in random order with interstimulus intervals from 400 to 550 ms. The loudness was approximately 70 dB SPL. The responses to 2-octave carrier stimuli for one subject are depicted here in detail, but all data for all subjects is analyzed below. The subjects reported normal hearing and no history of neurological disorder. The procedures were approved by the University of Maryland institutional review board and written informed consent was obtained from the participants.

Recordings were performed in a magnetically shielded room, using a 160-channel, whole-head axial gradiometer system (KIT, Kanazawa, Japan). The magnetic signals were bandpassed between 1 Hz and 200 Hz, notch filtered at 60 Hz, and sampled at 1000 Hz. All 157 neural channels were denoised with a Block-LMS adaptive filter using the 3 reference channels.

The measured responses from 50 to 1050 ms post-stimulus were concatenated, giving 12 total responses ($T$ = 100 s) for each channel. The discrete Fourier transform was applied to the concatenated data. The whole-head SSR is the magnitude and phase at the modulation frequency for each channel.

Pairs of dipoles sources were estimated using the complex Sarvas approximation described above and a modified simplex search (Uutela et al., 1998). 5 of the 36 frequency $\times$ bandwidth $\times$ subject searches did not lead to two separated dipoles and were discarded.

Calculations were performed in MATLAB (MathWorks, Natick, Massachusetts), which treats complex numbers transparently.

*2.6. Models and Simulations*

The complex field configuration due to a pair of dipoles, found from the complex Sarvas approximation to the complex data set shown in Fig. 1a., is shown in Fig. 3a. The parameters of that dipole pair, and of the dipole pairs analogously derived from the complex data sets shown in Fig. 1b-e, are given in Table I.

[Fig. 3 about here]

[Table I about here]

The complex magnetic field shown in Fig. 3a. is faithful to the most prominent features from in data shown in Fig. 1a: all peaks (regions of largest phasors) are in the same locations, with the same relative strengths, covering the same areas, and with phases in the same directions. The phases are not constant within each hemisphere, especially so in the right hemisphere. It will be seen below that this is due to non-zero sharpness.

Simulated complex magnetic fields were generated from pairs of ideal complex dipole point sources in left and right auditory cortex. The resulting complex fields are shown in Fig. 3b-e. To ease comparison with the experimental data shown in Fig. 1a and the dipole fit shown in Fig. 3a, the location, orientation, and intensity of every simulated dipole is set equal those of the pair of dipoles used in Fig. 3a, but the phase, sharpness and secondary orientation have been idealized: the phase is constant for both dipoles and across all simulations; the left dipole has sharpness $\eta = 0$ across all simulations; the right dipole has sharpness with the values (0.0, 0.25, 0.50, 1.0) with secondary orientation in the same direction.

The simulation with $\eta = 0$ in both hemispheres (Fig. 3b) has constant phase (mod $180°$) for all sensors. This is the single orientation approximation. It has a very simple phase structure, but it fares poorly in the right hemisphere at approximating the data in Fig 1a. The simulations with intermediate right hemisphere sharpness (Fig. 3b-c) show that slowly varying phase is generated only by a fully complex dipole (note that right hemisphere dipole in Fig. 3a has $\eta = 0.41$). The simulation with $\eta = 1$ in the right hemisphere has no preferred orientation, and the phase distribution in the magnetic field shows phases of all angles. Note that the nonzero $\eta$ cases show that phase structure of mild to high complexity is easy to generate even in the idealized case of zero noise.

## 3. Results

### *3.1. Transfer Function Example*

As an example of the complex equivalent-current dipole analysis method, we calculate a set of transfer functions: the response strength and phase of the complex equivalent-current dipole, as a function of the auditory stimulus modulation frequency. The transfer functions are calculated and compared for three carriers different bandwidths. The auditory whole-head SSR is measured for the four stimulus modulation frequencies, as shown for in Fig. 1b-e, and the response is characterized by the single, complex, equivalent-current dipole in each hemisphere (parameters summarized in Table I for one subject and one bandwidth). The response strength is measured by the dipole's $|\mathbf{v}_{\mathbf{Max}}|$, and its phase by the dipole's phase $\theta_{Max}$. The sharpness is ignored for this analysis. Separate transfer functions are calculated for each hemisphere.

The transfer functions, averaged over all subjects and both hemispheres, are illustrated in Fig. 4, with separate plots for amplitude and phase. Phases are unwrapped (from their $180°$ ambiguity) to be downwardly monotonic. Plotted separately are the averages over all subjects of the corresponding Right-minus-Left responses (dashed lines). The hemispheric differences in amplitude are small relative to their means. The hemispheric differences in phase are more noticeable; phase differences between the hemispheres imply a differential time lag in their processing (e.g. $45°$ at 32 Hz gives 4 ms difference).

Note that the stimulus frequencies chosen for this example, by omitting 40 Hz, miss much of the interesting behavior known to occur at that frequency (Ross et al., 2000; Ross et al., 2005).

In short, the complex dipole captures both the strength and the phase of a response in an unambiguous manner, without the need for ad-hoc methods otherwise used determine a single dipole origin from time varying signal.

[Fig. 4 about here]

Three subjects are not sufficient to draw conclusions (or calculate trustworthy confidence intervals) regarding any of the observations above, but it appears that bandwidth may not be an important parameter in the transfer functions for frequencies above 16 Hz.

*3.2. Noise Analysis of the Distribution of Sharpness Values*

Data corrupted by noise will show additional spatial phase variation over the noiseless case, and low spatial frequency spatial phase variation is likely to influence the complex dipole fits. This potentiality can be explored by plotting the sharpness as a function of noise. Here we estimate noise with the magnitude of the correlation coefficient defined in Eq. 6.

[Fig. 5 about here]

Fig. 5 shows sharpness as a function of correlation coefficient magnitude, with points identified by their stimulus frequency (a) or their stimulus bandwidth (b). First, we examine the data by stimulus frequency. Typical 16 Hz responses have the lowest correlation between the model and the data of any of the stimulus frequencies, and hence are the noisiest. Their sharpness values are widely distributed between 0 and 1, and the most parsimonious explanation is that those estimates of sharpness are contaminated by noise. In contrast, the responses at 48 and 64 Hz are striking in their higher correlation coefficient values, implying less corruption by noise. Comparing the two, it can be seen that for similarly high correlation coefficients, as a population the 48 Hz responses have sharpness values closer to zero than those of 64 Hz. As stated above, three subjects are not sufficient to draw conclusions, but it is plausible that responses at 48 Hz may be better approximated by the single orientation approximation than corresponding responses at 64 Hz. No such effects are seen as a function of stimulus bandwidth.

**4. Discussion**

The complex magnetic field distributions occurring from Fourier transformed MEG data have a natural interpretation as oscillations with a specified amplitude and phase. Visual representations of the complex responses over the whole head are invaluable in identifying structure and patterns in the whole head response. The addition of (real) magnetic field contours, derived from the complex field, increase a viewer’s ability to see natural structures such as dipolar configurations.

Using the complex generalization of the spherical head model, we can find complex equivalent-current dipoles that are the best fit to the whole-head complex magnetic field. In addition to its location, a complex equivalent-current dipole vector has six degrees of freedom, twice that of a real vector: a strength and orientation (similar to all the degrees of freedom of a real vector), a complex phase, and two additional parameters—the sharpness and a secondary orientation.

Isolated neural current sources are well described by the single orientation approximation ($\eta = 0$) and behave similarly to a real dipole but with the addition of phase. In contrast, closely spaced discrete sources with differing orientations produce an effective complex dipole with non-zero sharpness and a conspicuous secondary orientation. Thus, any complex dipole of moderate sharpness constitutes evidence for the existence of multiple sources. The use of this technique to reveal multiple sources is less susceptible to error than an explicit multiple source fit because fitting to closely-spaced sources is prone to error, requires more parameters than a single complex dipole fit, and may be genuinely unattainable with the limited spatial resolution of MEG (Lutkenhoner, 2003).

The presence of closely spaced, difficult to separate, sources can be recognized by detecting transitions from high sharpness to low (or vice versa). One example illustrated above arises in the search for SSR sources as a function of modulation frequency or carrier bandwidth. The former search is motivated by group delay evidence that high and low frequency SSR responses originate from different sources (Ross et al., 2000; Schoonhoven et al., 2003). Another example might be the analogous search for narrowband SSR sources as a function of carrier frequency (tonotopy). In a different example, the change in the number of closely spaced neural sources, as a function of stimulus parameter, might reflect different states in an experiment designed to detect neural correlates of attention.

Historically, Lehmann and Michel (1989; 1990) described a process for determining a complex dipole source from complex EEG data, but the process explicitly requires that the dipole be fit to data with a single phase. This is equivalent to requiring the single orientation approximation  (illustrated in Fig. 3b) and does not allow for all six degrees of the complex source. Lutkenhoner (1992) went substantially further and showed that standard MEG localization methods generalize straightforwardly to complex data and naturally result in complex neural sources. Fully complex sources using all six degrees of freedom, however, are not considered. Indeed, all the illustrative examples are forward model simulations with single orientation.

Finally, since the use of fully complex sources is an *analysis* method, it is straight-forward to apply it to previously obtained (periodic or oscillatory) data as well as to new experiments. Applications range from using the complex dipole to capture both the strength and the phase of a response in an unambiguous manner, to explicit analyses of the dipole sharpness as a measure of neural source configuration.

**Acknowledgements**

We thank David Poeppel and Kensuke Sekihara for help and discussions. This work was supported by a UMCP GRB award to J.Z.S. and by NIH grant R01-DC05660 to David Poeppel for Y.W..

| **Parameter** | | 16 Hz | 32 Hz | 48 Hz | 64 Hz |
|---|---|---|---|---|---|
| Amplitude | (Left) | 28 dB | 32 dB | 25 dB | 15 dB |
| $\lvert \mathbf{v}_{\mathbf{Max}} \rvert$ | (Right) | 27 dB | 33 dB | 25 dB | 15 dB |
| Phase | (Left) | 107º | 30º | -45º | -107º |
| $\theta_{Max}$ | (Right) | 116º | 25º | -46º | -109º |
| Sharpness | (Left) | 0.26 | 0.27 | 0.07 | 0.21 |
| $\eta$ | (Right) | 0.17 | 0.41 | 0.11 | 0.03 |
| Location *x* *y* *z* | (Left) | 45 mm<br>17 mm<br>-4 mm | 35 mm<br>7 mm<br>17 mm | 36 mm<br>14 mm<br>21 mm | 19 mm<br>12 mm<br>28 mm |
| *x* *y* *z* | (Right) | -52 mm<br>3 mm<br>13 mm | –34 mm<br>8 mm<br>-8 mm | –46 mm<br>19 mm<br>13 mm | –39 mm<br>12 mm<br>18 mm |
| Orientation θ φ | (Left) | 31º<br>300º | 50º<br>258º | 140º<br>71º | 123º<br>357º |
| θ φ | (Right) | 14º<br>6º | 29º<br>233º | 161º<br>119º | 143º<br>215º |
| GOF | | 0.51 | 0.70 | 0.52 | 0.45 |
| $\angle r$ | | 7º | –9º | 16º | -3º |
| $\lvert r \rvert$ | | 0.81 | 0.86 | 0.89 | 0.84 |

Table I. Complex Dipoles. The dipole (left and right) and evaluation (whole head) parameters, for dipole fits to the data illustrated in Fig. 1. The orientation parameters θ and φ refer to elevation (downward from the *z*-axis) and azimuth. The secondary orientation is omitted since the Sarvas model requires it to be radial.

Fig. 1. Whole Head Complex MEG. The whole-head complex SSR from one subject in an auditory MEG experiment. The 157 channels are shown on the surface of a flattened head. Each arrow represents the complex field value at a sensor. (a) The whole head SSR for a 2-octave broadband stimulus, amplitude modulated at 32 Hz. Each hemisphere is dominated by a classic pattern of dipole-like generated activity, but in this case, the field is complex. (Inset) The same whole-head complex SSR but without the contour map, making the dipolar patterns much harder to discern. (b-e) Responses from the same subject and carrier for four modulation frequencies: 16 Hz, 32 Hz (also shown twice in a), 48 Hz, and 64 Hz. In every case, both hemispheres are dominated by a classic pattern of dual-dipole-like generated activity, with variation in location, size, and strength across stimuli. Phasor arrows in all four examples are all scaled to the same (arbitrary) strength. Contour map colors are scaled individually to emphasize their patterns. Subject R0292.

Fig. 2. Ellipse Swept by a Complex Vector. The ellipse swept out by a complex vector as phase (or time) increases throughout an entire cycle. At the start of the cycle, the vector is equal to $\mathbf{v}_{\mathbf{Re}}$, changing direction and length until it is equal to $\mathbf{v}_{\mathbf{Im}}$ after one quarter cycle, and then continuing around the ellipse. When the phase has advanced $\theta_{Max}$, the length of the vector is at its maximum, corresponding to the semimajor axis $\mathbf{v}_{\mathbf{Max}}$. When the phase has advanced to $\theta_{Min} = \theta_{Max} + 90°$, the length of the vector is at its minimum, corresponding to the semiminor axis $\mathbf{v}_{\mathbf{Min}}$. Note that the portrayed angles are *phase* angles, not spatial angles.

Fig. 3. Model Fit and Simulations. The whole-head complex SSR from model-fit and simulated auditory MEG experiments. The complex magnetic field is generated by a pair of complex point dipole sources. (a) The complex magnetic field generated by the pair of complex point dipole sources fit to the data illustrated in Fig. 1a, using the complex Sarvas model. (b-e) The location, orientation, and intensity of every simulated dipole is set equal those of the pair of dipoles used in (a), but the phase, sharpness and secondary orientation have been idealized: the phase is constant for both dipoles and across all simulations and the left dipole has sharpness $\eta = 0$ across all simulations. (b) The right hemisphere dipole has $\eta = 0$ as well. Each hemisphere is dominated by a classic pattern of dipole-like generated activity, and, in this case, the phase of the complex field is constant (mod $180°$) everywhere. (c) The right hemisphere dipole gains a secondary orientation contribution with relative strength $\eta = 0.25$. The magnetic field in the right hemisphere is no longer constant phase, but has phase shifts of up to $90°$ for channels further from magnetic dipole peaks. The magnetic field in the left hemisphere is largely unaffected. (d) The right hemisphere dipole has $\eta = 0.5$. Over the right hemisphere, the phase shift for the medial channels is now substantial, and even some left hemisphere channels are affected in phase. There is a visual impression of phase flow. (e). The right hemisphere dipole has $\eta = 1$. Over the right hemisphere, channels with phase shifts of $90°$ can dominate over the original phase. The effect of on the medial and posterior left hemisphere is substantial, and the visual impression of phase flow is striking.

Fig. 4. Transfer Functions. Transfer functions derived from equivalent-current dipoles fit to each hemisphere averaged over all subjects. (a) Amplitude in dB as a function of stimulus frequency for each carrier bandwidth. Mean amplitude over hemispheres (solid lines); Right-minus-Left amplitude difference (dashed lines). (b) Phase in degrees as a function of stimulus frequency for each carrier bandwidth (using circular mean). Mean phase over hemispheres (solid lines); Right-minus-Left phase difference (dashed lines).

Fig. 5. Noise Analysis for Sharpness Distribution. Sharpness as a function of correlation coefficient magnitude, with points identified by their stimulus frequency (a) and their stimulus bandwidth (b). Probability of the neural source being a compound source increases upward. Inferred reliability increases to the right.

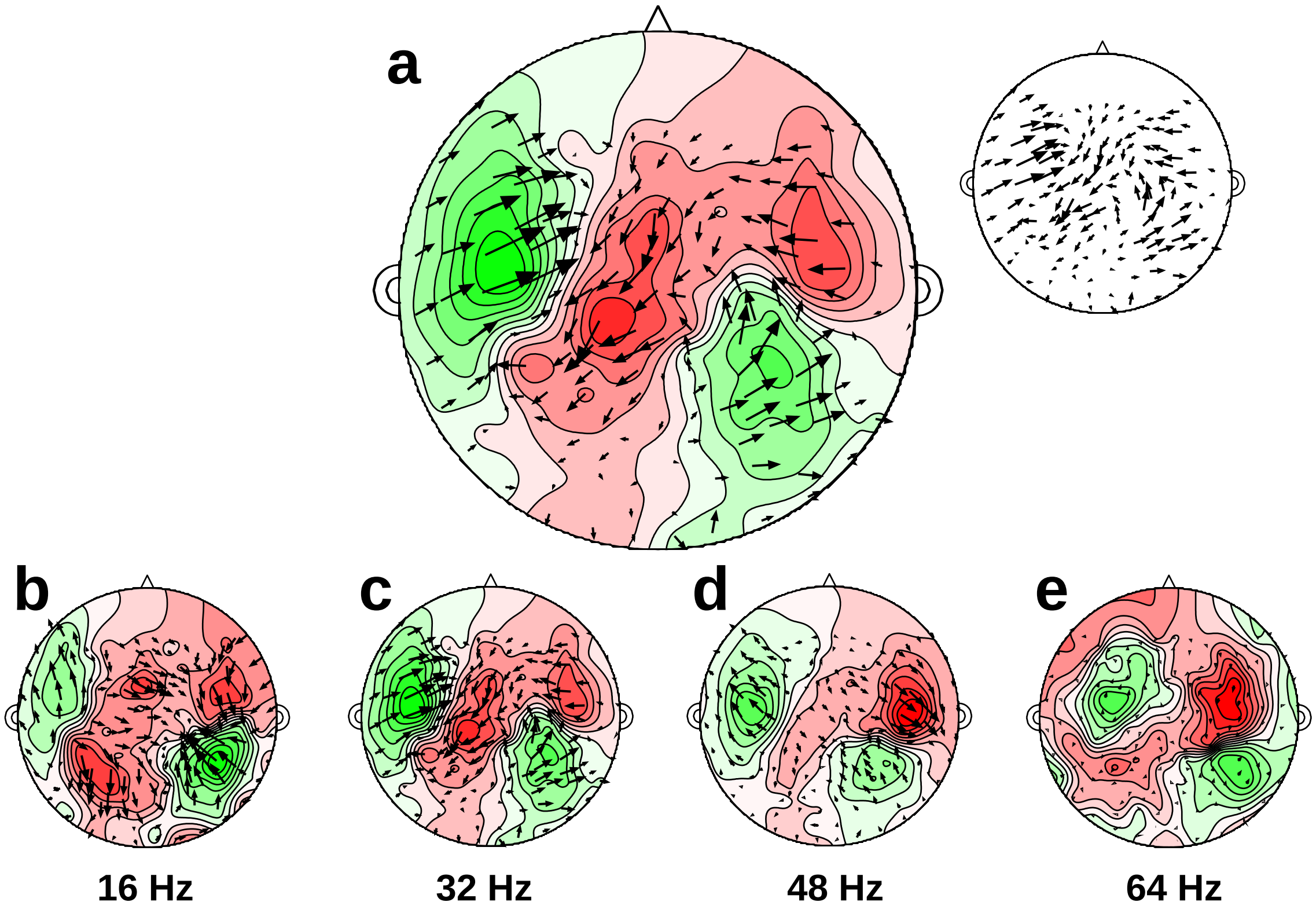

Figure 1

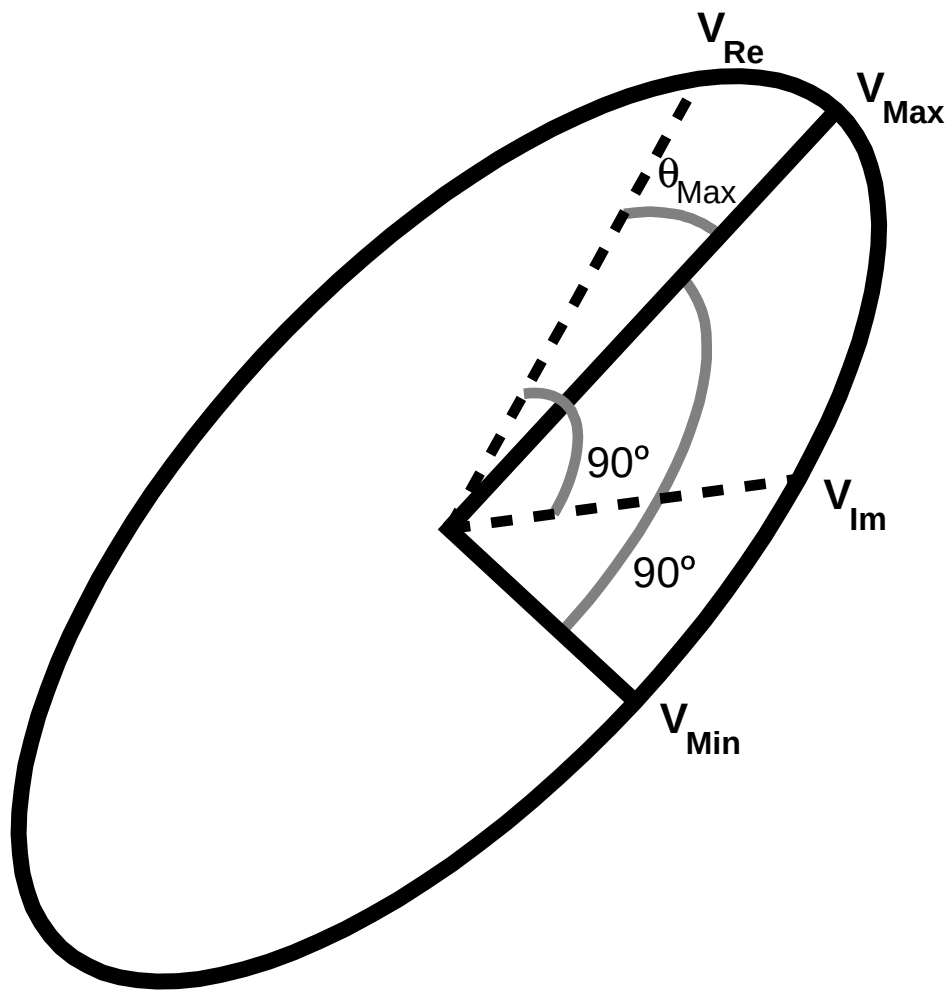

Figure 2

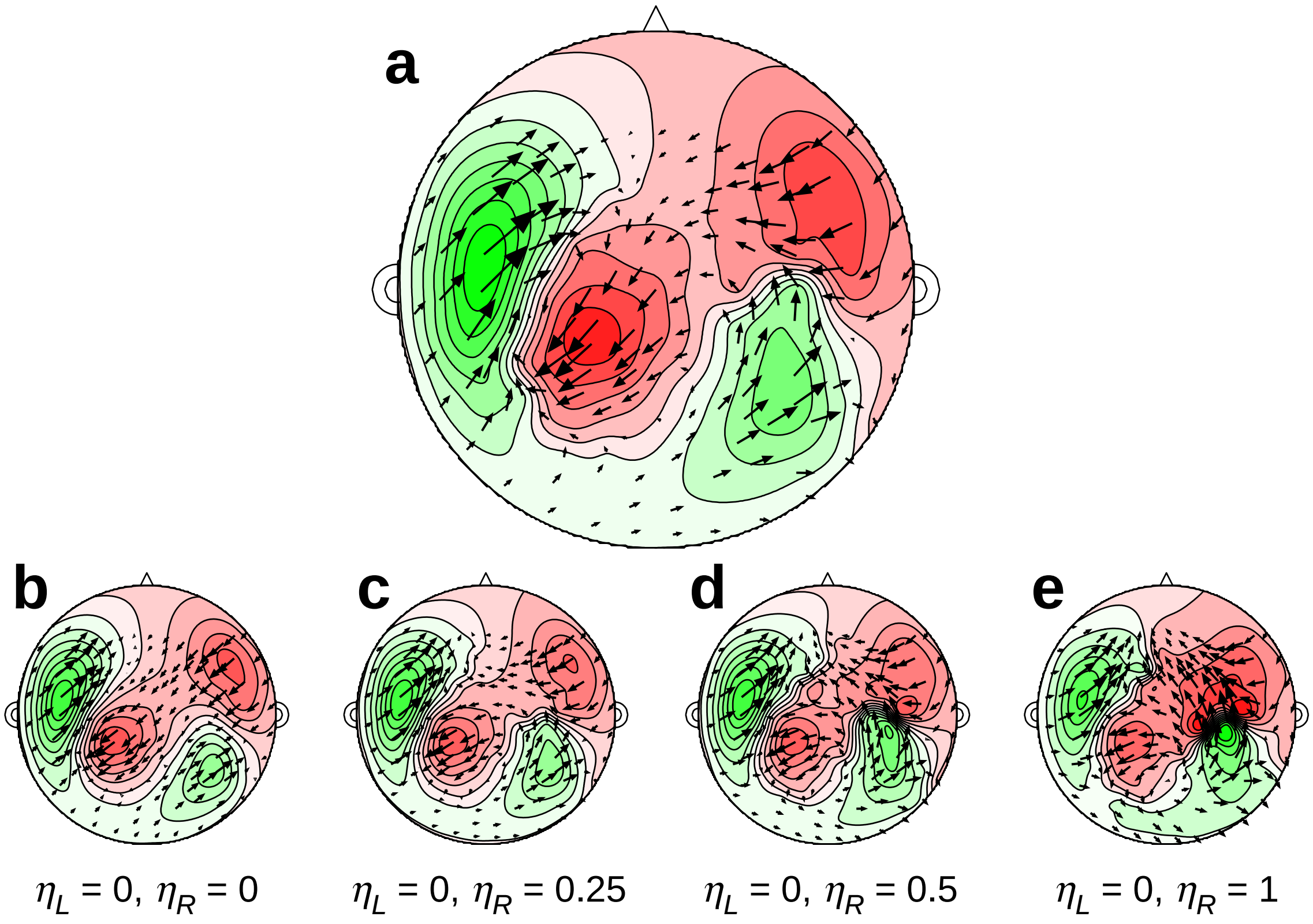

Figure 3

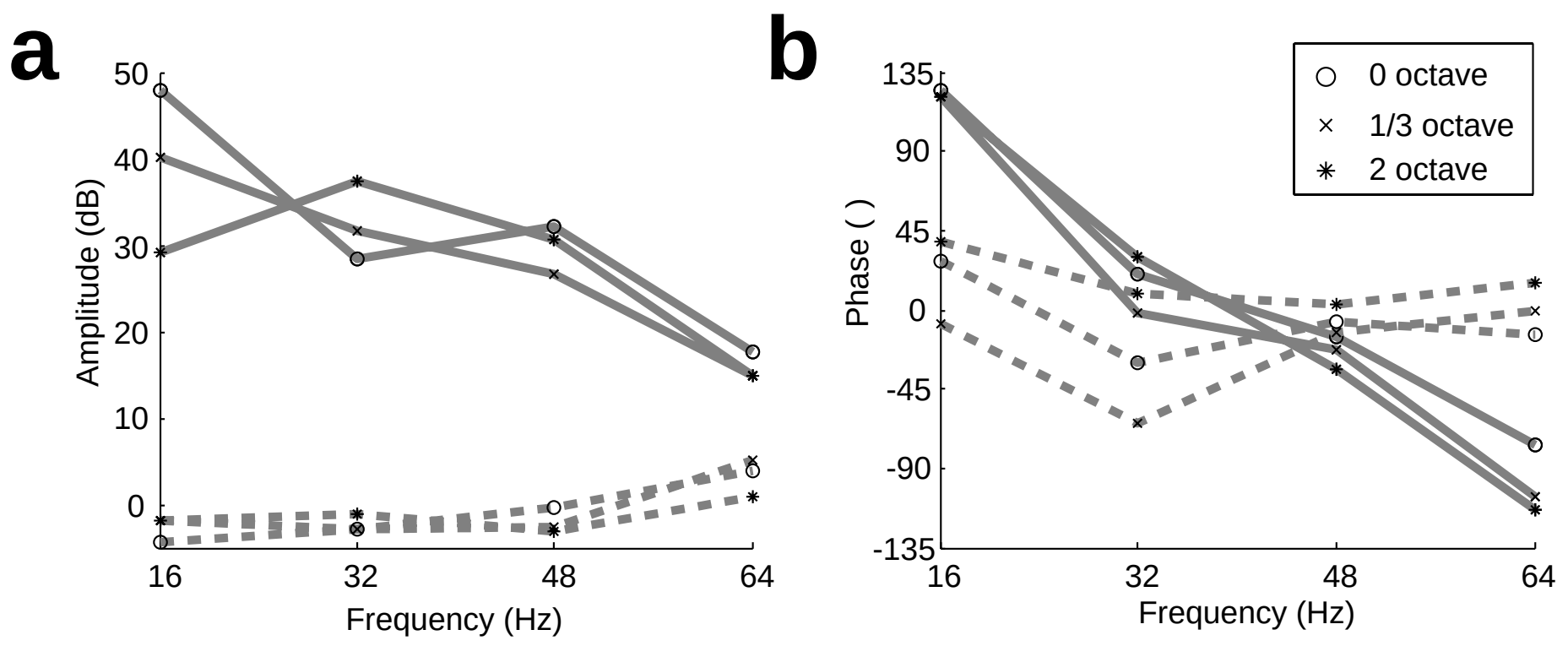

Figure 4

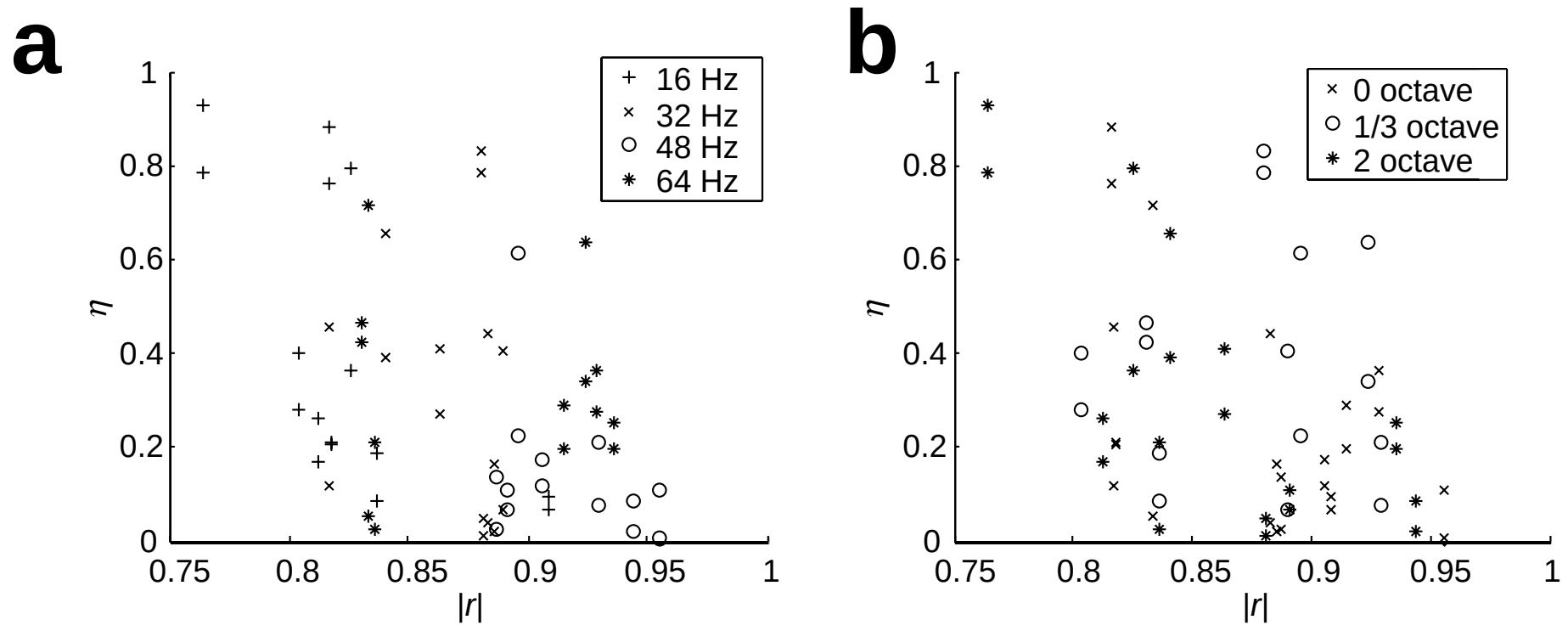

a
b
16 Hz
32 Hz
48 Hz
64 Hz
0 octave
1/3 octave
2 octave
η
|r|
0
0.2
0.4
0.6
0.8
1
0.75
0.8
0.85
0.9
0.95

Figure 5